\documentclass[%
 reprint,
superscriptaddress,
nofootinbib,
bibnotes,
 amsmath,amssymb,
 aps,
]{revtex4-1}

\usepackage{graphicx}
\usepackage{dcolumn}
\usepackage{braket}
\usepackage{gensymb}
\usepackage{mathtools}
\usepackage{bm}
\usepackage{amsmath}
\usepackage{float}
\usepackage{color}
\usepackage{units}
\usepackage{lipsum}
\usepackage{csquotes}
\usepackage[makeroom]{cancel}
\usepackage{multirow}
\usepackage{booktabs}
\usepackage{amssymb}
\usepackage{color}

\usepackage{hyperref}

\makeatletter
\newcommand*{\rom}[1]{\expandafter\@slowromancap\romannumeral #1@}
\makeatother

\begin{document}

\preprint{APS/123-QED}

\title{Post-Selection and Counterfactual Communication}
\author{David R. M. Arvidsson-Shukur}
\affiliation{%
Department of Mechanical Engineering, Massachusetts Institute of Technology, Cambridge, Massachusetts 02139, USA}
\affiliation{%
 Cavendish Laboratory, Department of Physics, University of Cambridge, Cambridge CB3 0HE, United Kingdom
}%

\author{Crispin H. W. Barnes}
\affiliation{%
 Cavendish Laboratory, Department of Physics, University of Cambridge, Cambridge CB3 0HE, United Kingdom
}%

\date{\today}

\begin{abstract}
A recently proposed quantum protocol for counterfactual communication [Y. Aharonov and L. Vaidman, Phys. Rev. A 99, 010103(R), 2019] relies on post-selection to eliminate the weak trace in the transmission channel. We show that the post-selection in this protocol additionally eliminates the flow of Fisher information from transmitter to receiver. However, we also show that a classical communication protocol with post-selection can be counterfactual. Hence, we argue that post-selection should not be allowed in genuine quantum counterfactual communication. In the proposed quantum protocol, the probability of discarding an event by post-selection tends to zero with an increasing number of ideal optical components. But the counterfactual violation strength tends to infinity at a faster rate. Consequently, the quantum protocol is not counterfactual proper.
\end{abstract}

\maketitle


\section{Introduction}

In standard communication protocols, a message is carried by signal particles propagating from the information transmitter to the information receiver \cite{Shannon48}. Recent years have seen an intense debate about whether or not quantum mechanics can allow a message to be sent without the transmitter sending any particles to the receiver---that is, \textit{counterfactually} \cite{Salih13, Gisin13, Vaidman14, Salih14, Li15, Vaidman15, ArvShukur16, Vaidman16, Li16, Griffiths16, ArvShukur17-2, Calafell18}.

A protocol, first proposed by Salih \textit{et al.}, is based on a nested concatenated Mach-Zehnder interferometer (MZI) and the quantum Zeno effect \cite{Salih13}. Superficially this protocol achieves so-called type-\rom{1} counterfactual communication (CFC): the quantum celebrities Alice (the information receiver) and Bob (the information transmitter) can communicate without any particles crossing the transmission channel that separates them. This protocol has been subject to intense criticism. First, the protocol requires an interferometer with tens of thousands of optical components to achieve high ($> 95 \, \%$) success probabilities \cite{ArvShukur16}. Second, an analysis of the flow of Fisher information shows that counterfactuality is satisfied only in the presence of perfectly noiseless quantum channels \cite{ArvShukur17-2}. Third, Alice's particles leave a \textit{weak trace} in Bob's laboratory, which according to several scholars means that the particles crossed the transmission channel \cite{Vaidman13, Danan13, Vaidman14, Aharanov17}. Fourth, unless an infinite number of \textit{perfect} optical components is used, post-selection has to be employed in order to discard events that violate counterfactuality \cite{Calafell18}. 

In their innovative article, Ref. \cite{Aharonov19}, Aharonov and Vaidman (AV) modify the counterfactual communication protocol of Salih \textit{et al.} \cite{Salih13}, claiming to make it robust against the weak-trace argument. Their interferometer protocol has two processes, by which Bob can send a $0$-bit or a $1$-bit to Alice, with some probability of success. If such a process succeeds, the particle that Alice detects left no weak trace \cite{Vaidman13} in Bob's part of the interferometer, meaning that Bob can not infer the particle's presence via a local weak measurement.

 In this paper we use methods from information theory to study the protocol of AV. In particular we study the flow of Fisher information in the interferometers used by AV. We find that in successful bit-transmission processes, the protocol particles carry no Fisher information about Bob's laboratory to Alice. From a Fisher information perspective, the AV protocol is counterfactual if post-selected on successful bit-transmissions. However, we argue that this is not enough to comply with the original type-\rom{1} definition of CFC, which does not consider any post-selection requirement \cite{Salih13}. We design a classical protocol, which can achieve CFC with post-selection, and conclude that  post-selection is undesirable if CFC is to describe a non-classical phenomenon. We then show that the events, which are discarded by post-selection in the AV protocol, violate counterfactuality to a level where the protocol cannot be called counterfactual proper.

\section{Fisher-information measure of counterfactuality}

To evaluate the level of counterfactuality of an experiment we need a measure of inter-measurement particle presence. One conceptually beautiful suggestion of such a measure is the weak trace \cite{Vaidman13}. However, in the words of Vaidman, \textit{et al.}: `The weak value requires the two-state vector formalism for its definition' \cite{Vaidman17}; and the two-state vector formalism is a controversial interpretation of quantum mechanics, which has generated criticism of the weak-trace measure \cite{Svensson13, Li13, Li16, Hashmi16}.

 In Ref. \cite{ArvShukur17-2} we (together with A. Gottfries) suggested how the flow of Fisher information can be used to study the inter-measurement presence of a quantum particle, from an interpretation-independent and operational perspective. The Fisher information about an unknown parameter, $\theta$, is 
 \begin{equation}
 F(\theta)=\sum_{i} p_i(\theta) [\partial_{\theta} \ln{p_i(\theta)}]^2 ,
\end{equation}   where $p_i(\theta)$  is the probability of measurement outcome $i$ given $\theta$ \cite{bCover06}. 

Consider a horizontally polarised photon travelling through a non-polarising interferometer. Perfectly noiseless quantum channels are purely fictional, and we introduce a \textit{vanishingly weak} polarisation rotation, a ``tagging'', to mimic an unwanted polarisation disturbance \cite{ArvShukur17}. The tagging is the only component acting on the polarisation degree of freedom in the interferometer. We thus argue that output photons, which carry polarisation-encoded information about the tagging, have been present at the location of the tagging. This toy model can be used to study how information flows through interferometers. We quantify the \textit{extent} to which a particle has been present at the location of the tagging by $\mathcal{D}_{vio} = F(\theta)/F_{ref}$, where $\theta$ is the parameter that sets the tagging interaction \cite{ArvShukur17-2}. $F_{ref}$ is the reference Fisher information obtained had the whole particle wavepacket propagated freely through the location of the interaction. $\mathcal{D}_{vio} \geq 1$ reveals a full counterfactual violation.

Here we analyse interferometers where the particle can move in and out of Bob's laboratory several times. Every time the particle does so,  type-\rom{1} counterfactuality is violated. We extend the Fisher-information measure of the counterfactual violation strength from Ref. \cite{ArvShukur17-2} to the scenario of multiple paths through Bob's laboratory:
\begin{equation}
\mathcal{D}_{vio} = n_{\gamma} \lim_{\overline{\theta} \rightarrow 0} \sum_{t=1}^{N_T} \frac{ F^{0}(\theta_t) + F^{1}(\theta_t) }{2 F_{ref}} ,
\end{equation}
where $n_{\gamma}$ is the average number of interferometer evaluations per transmitted bit, and $\overline{\theta} \equiv (\theta_{1}, \, ..., \, \theta_{N_T})$ is a vector containing all the parameters that set the $N_T$ tagging interactions placed at each point where the protocol particle enters or re-enters Bob's laboratory. $F^{0}$ and $F^{1}$ are the Fisher information in a $0$-bit and a $1$-bit process, respectively, calculated from the output probabilities of the interferometer. We assume messages with a balanced number of $0$- and $1$-bits. In experiments without post-selection, our measure is directly proportional to the average sum of the integrated probability density that travels in and out of Bob's laboratory per transmitted bit (in the Schr\"odinger picture).

\section{A classical CFC protocol with post-selection}
In type-\rom{1} CFC, a message is sent without any particles crossing the transmission channel between Alice and Bob \cite{Salih13, ArvShukur17-2}. Below we show that a classical protocol can transmit two bit values  counterfactually if post-selection is allowed.

Alice and Bob reside in respective laboratories, connected with a pipe, through which Bob can roll balls to Alice. The communication protocol is as follows
\begin{itemize}
\item{ The message is sent by Bob rolling balls to Alice. One ball, and bit, is transmitted per minute. $T_r < 1 \, \textrm{min}$, where $T_r$ is the time it takes one ball to roll from Bob to Alice.}
\item{On even minutes, if Bob wishes to transmit a $1$-bit ($0$-bit)  he rolls (does not roll) a ball through the pipe to Alice.}
\item{On odd minutes, if Bob wishes to transmit a $0$-bit ($1$-bit) he rolls (does not roll) a ball through the pipe to Alice.}
\end{itemize}

At even (odd) minutes, Alice records the value of the $0$-bit ($1$-bit)  without receiving Bob's protocol particles (balls), and without these particles crossing the transmission channel (the pipe). These bit values satisfy the type-\rom{1} definition of CFC. Post-selecting on Alice not receiving any particles, she receives all message bits counterfactually. We thus conclude that CFC with post-selection is not an inherent quantum phenomenon. A definition of non-classical CFC should require that the counterfactuality stems from quantum phenomena, rather than classical post-selection.

\begin{figure}
\includegraphics[scale=0.34]{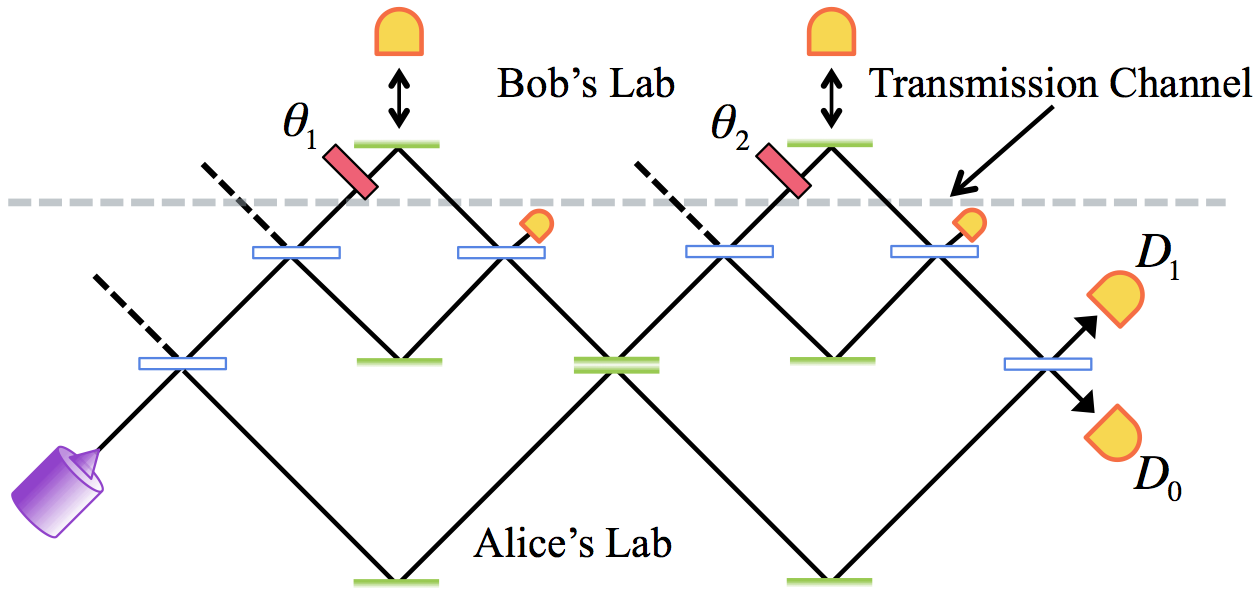}
\caption{The doubly nested Mach-Zehnder interferometer as used in the reduced CFC protocol of Ref. \cite{Aharonov19}. Solid and hollow rectangles represent mirrors and beam-splitters, respectively. Bob can insert mirrors ($0$-bit process) or detectors ($1$-bit process) in his laboratory. Vanishingly weak polarisation rotations have been inserted at both locations where Alice's photon can enter Bob's laboratory. The detectors detect single photons and their polarization. }
\label{fig:Double_NMZI}
\end{figure}

\section{Reduced AV CFC Protocol}

Vaidman and Aharonov first investigate a reduced CFC protocol, based on a doubly nested MZI. See Fig. \ref{fig:Double_NMZI}. The outer and inner beam-splitters have transmission probabilities of $4/5$ and $1/2$, respectively. Bob can insert mirrors or detectors in the upper path of the interferometer. The inner MZIs are tuned to give completely destructive interference towards their lower output paths when the mirrors are inserted. Both bit-transmission processes start by Alice sending a photon into the interferometer's lower path. To send a $0$-bit, Bob inserts his mirrors. To send a $1$-bit, Bob replaces his mirrors with detectors. Only the $0$-bit process can probabilistically trigger detector $D_0$. The tagging interactions rotate the polarisation state: $\ket{H}\rightarrow \cos(\theta)\ket{H} + \sin(\theta)\ket{V}$ and $\ket{V}\rightarrow \cos(\theta)\ket{V} - \sin(\theta)\ket{H}$, with $\theta \ll 1$. The reference Fisher information is  $F_{ref}=4$.

To satisfy counterfactuality, the reduced AV protocol requires the post-selection of only the events that trigger detectors $D_0$ or $D_1$. Calculating the Fisher information from the post-selected outcome probabilities, we find that $F(\theta_1)$ and $F(\theta_2)$ are strictly zero in the $1$-bit process and approaches zero as $\theta_1, \, \theta_2 \rightarrow 0$ in the $0$-bit process. From a post-selected perspective, the counterfactual violation strength is indeed approaching zero for both bit values. However, as we argued above, post-selection is undesirable in CFC. 

We thus proceed by calculating $F(\theta_1)$ and $F(\theta_2)$ with respect to the measurement statistics of all the detectors of the device, without post-selection. We find that $F^{0}(\theta_1)=\frac{8}{5}$ and $F^{0}(\theta_2)=\frac{4}{5}[1-\cos(\theta_1)]$ in the $0$-bit process, and $F^{1}(\theta_1)=\frac{8}{5}$ and $F^{1}(\theta_2)=\frac{2}{5}$ in the $1$-bit process.  The value of the type-\rom{1} counterfactual violation strength is then  $\mathcal{D}_{vio}=n_{\gamma}\frac{9}{20}$. 

Finally, let us estimate $n_{\gamma}$.  In Ref. \cite{Aharonov19}, the bit values are deduced from the statistics of detector $D_0$ only. For a successful $0$-bit transmission, $D_0$ must be triggered. In one $0$-bit process the probability for this is $P_{D_0}=1/25$. For the average $0$-bit error to be less than $5 \, \%$, we repeat each process $n_{\gamma}>74$ times. This gives a counterfactual violation strength of $\mathcal{D}_{vio} > 33.3$; the non-post-selected reduced protocol is not counterfactual.

\section{Full AV CFC Protocol}

By increasing the number of perfect optical components it is possible to improve the probability of successful bit-transmission in the AV CFC protocol \cite{Salih13, Aharonov19}. The extended interferometer is similar to that in Fig. \ref{fig:Double_NMZI}, but concatenates $N$ beam-splitters in the outer MZI, and  concatenates $M$ beam-splitters in each of the inner MZIs \cite{Aharonov19}. The outer and inner beam-splitters have transmission probabilities $\sin^2(\pi / 2 N)$ and $\sin^2(\pi / 2 M)$, respectively. To transmit bits with high success probabilities the protocol requires that $M \gg N \gg 1$. For perfect optical components, and if $M,\; N \rightarrow \infty$, the success probability of both bit processes approaches unity, and $n_{\gamma}=1$. This is an unrealistic protocol, and not ``a very reliable communication protocol'' as suggested in Ref. \cite{Aharonov19}.


We consider vanishingly weak tagging interactions at each particle path entering or re-entering Bob's part of the interferometer. As in the reduced AV protocol, the Fisher information, and thus also counterfactual violation strength, approaches zero when we post-select on outcomes at $D_0$ and $D_1$. But, does the counterfactuality of the AV CFC protocol stem from a genuine \textit{quantum} phenomenon, or simply from the post-selection of specific outcomes? 

Let us analyse the full AV CFC protocol without post-selection. For the $1$-bit process, the counterfactual violation is equal to the probability of detection in Bob's laboratory, which approaches zero for large numbers of beam-splitters. For the $0$-bit process the counterfactual violation  is
\begin{align}
\mathcal{D}_{vio}^0 = & \sum_{n=1}^{N-1}\cos^{2(n-1)}\Big( \frac{\pi}{2N} \Big) \sin^{2}\Big( \frac{\pi}{2N} \Big) \nonumber \\
& \times \sum_{m=1}^{M-1} \sin^{2}\Big( \frac{m \pi}{2M} \Big) .
\end{align}
For large values of $N$ and $M$ this simplifies significantly:
\begin{align}
\mathcal{D}_{vio}^0 \approx & \frac{\pi^2}{4N} \times \frac{M}{2} ,
\end{align}
such that $\mathcal{D}_{vio}^0 \gg 1$ for $M \gg N \gg 1$. We recall that if $\mathcal{D}_{vio} > 1$ the protocol is not counterfactual. Consequently, the AV CFC protocol is invalid without post-selection, even if $M, \; N \rightarrow \infty$. An analysis of the weak trace in the events discarded by post-selection leads to a similar conclusion. We can thus conclude that the quantum interferometer used in the AV CFC protocol does not eliminate counterfactual violations. This elimination is acquired instead by deploying post-selection---as in the classical CFC protocol above. 

\section{Discussion}

Aharonov and Vaidman write: ``The possibility to communicate between spatially separated regions, without even a single photon passing between the two parties, is an amazing quantum phenomenon'' \cite{Aharonov19}. Here, we have shown that CFC with post-selection can be achieved classically too, without any ``quantum phenomenon''. Hence, we argue that a truly counterfactual quantum experiment should operate without post-selection. The AV CFC protocol fails this requirement.\footnote{The same is true for the post-selected CFC protocol suggested in Ref. \cite{Salih18}. }

The AV protocol is nevertheless interesting. The post-selection in this protocol is different than in our classical protocol. In the classical protocol the probability of discarding a non-counterfactual event is $1/2$. In the AV protocol, the rate of discarding message bits by post-selection decreases with an increasing number of optical components. The surprising result is that the counterfactual violations can be distilled to a small number of events.\footnote{See Ref. \cite{ArvShukur19} for an outline of how post-selection can distill Fisher information beyond classical limits.} However, with the same increase of optical components, the counterfactual violation strength of these events tends to infinity at a faster rate making the AV protocol violate counterfactuality if post-section is prohibited.

Finally, we would like to direct the reader's attention to our own definition of CFC: the type-\rom{2} definition \cite{ArvShukur16, ArvShukur17-2}. This definition also considers a single-particle protocol, but relaxes slightly the restrictions on the particle propagation. Particles are allowed to travel in the opposite direction to the message, from information receiver to information transmitter. But particles are not allowed to travel from the  transmitter to the receiver. In our type-\rom{2} CFC protocol information and particles counterpropagate. A recent work, Ref. \cite{Calafell18}, experimentally demonstrate this non-classical communication protocol, without post-selection, and with a success rate of $ \sim 99 \, \%$. 

\vspace{1mm}

The authors would like to thank Irati Alonso Calafell for useful discussions. D.R.M.A.-S. acknowledges support from the EPSRC, the Sweden-America Foundation, and the Lars Hierta Memorial Foundation.

\bibliography{CFC2Comment}

\end{document}